\documentstyle[twocolumn,epsf,aps,prb,floats]{revtex}
\begin{document}
\draft

%%2col
\twocolumn[\hsize\textwidth\columnwidth\hsize\csname @twocolumnfalse\endcsname
%% start of wide text
%%2col

\title{Theory of Scanning Tunneling Spectroscopy of a
       Magnetic Adatom on a Metallic Surface}
\smallskip
\author{Avraham Schiller$^{1}$ and Selman Hershfield$^{2}$}
\address{
         $^1$ Racah Institute of Physics, The Hebrew University,
              Jerusalem 91904, Israel\\
         $^2$ Department of Physics and National High Magnetic
              Field Laboratory, University of Florida,\\
              PO Box 118440, Gainesville, FL 32611}
\date{\today}
\maketitle
\smallskip

\begin{abstract}
A comprehensive theory is presented for the voltage, temperature,
and spatial dependence of the tunneling current between a scanning
tunneling microscope (STM) tip and a metallic surface with an
individual magnetic adatom. Modeling the adatom by a nondegenerate
Anderson impurity, a general expression is derived for a weak
tunneling current in terms of the dressed impurity Green function,
the impurity-free surface Green function, and the tunneling matrix
elements. This generalizes Fano's analysis to the interacting case.
The differential-conductance
lineshapes seen in recent STM experiments with the tip directly
over the magnetic adatom are reproduced within our model, as is the
rapid decay, $\sim 10\AA$, of the low-bias structure as one moves
the tip away from the adatom. With our simple model for the
electronic structure of the surface, there is no dip in the
differential conductance at approximately one lattice spacing
from the magnetic adatom, but rather we see a resonant enhancement.
The formalism for tunneling into small clusters of magnetic
adatoms is developed.
\end{abstract}

\smallskip
%% 72.10.Fk -- Scattering by point defects, dislocations, surfaces,
%%             and other imperfections (including Kondo effect)
%% 72.15.Qm -- Scattering mechanisms and Kondo effect
%% 61.16.Ch -- Scanning probe microscopy: scanning tunneling, atomic
%%             force, scanning optical, magnetic force, etc.
\smallskip

%%2col
%% end of wide text
]
\narrowtext
%%2col

\section{Introduction}
Recently, a scanning tunneling microscope (STM) was used in two
separate experiments~\cite{STM98_ce,STM98_co} to directly probe
the local electronic structure of an isolated magnetic adatom on
a metallic surface. By measuring the tunneling current from the
substrate to the STM tip, a narrow resonant feature was seen in
the differential conductance when the tip was placed directly
above the magnetic adatom: an antiresonance in the case of a
cerium adatom on the (111) surface of silver,~\cite{STM98_ce}
and an asymmetric Fano resonance in the case of a cobalt
adatom on the (111) surface of gold.~\cite{STM98_co}
These resonant structures gradually disappeared upon removing
the tip from the adatom. Specifically, the asymmetric lineshape
for Co on Au(111) first evolved into a more symmetric dip,
before disappearing altogether at a lateral distance of about
12\AA\ from the Co adatom.~\cite{STM98_co} Both experiments were
interpreted as a manifestation of the Kondo resonance that
develops due to the screening of the local moment on the
magnetic adatom by the substrate conduction electrons.

While similar observations of the Kondo effect for a single
magnetic impurity were recently reported both in quantum
dots~\cite{QD1,QD2} and in metallic point contacts,~\cite{RB94}
STM spectroscopy offers the unique ability to spatially resolve
the electronic
structure around the magnetic adatom. This provides direct
information about the screening of the local moment, allowing
for critical comparison between theory and experiment.
Another novel aspect of STM spectroscopy in this context is
the quantum-mechanical interference between direct tunneling
into the magnetic adatom and tunneling into the underlying substrate
conduction electrons, as first discussed by Fano~\cite{Fano61}
for the noninteracting case. For a quantum dot placed in between
two metallic leads, the analogous interference is between
Kondo-assisted tunneling through the dot and direct tunneling
between the leads, the latter being extremely small.

Motivated by the recent STM experiments, this paper provides
a comprehensive theory for the voltage, temperature, and spatial
dependence of the tunneling current between an STM tip and a
metallic surface with an individual magnetic adatom. Modeling
the adatom by a nondegenerate Anderson impurity, a general
expression is derived for a weak tunneling current in terms of the
fully dressed impurity Green function and the impurity-free
surface Green function. The impurity Green function is
evaluated in turn using the non-crossing approximation (NCA),
while the surface Green function is obtained within a
tight-binding model with free boundary condition at the
surface. This allows for a consistent description of the
energy and the spatial dependence of the tunneling current,
as is required in this case. Both cases of point tunneling
and that of a finite spatial extent in the tunneling matrix
element between the tip and the substrate conduction electrons
are considered within this model.

As expected of the Kondo effect, a sharp resonant structure is
found to develop in the low-voltage differential conductance as
the temperature is lowered down to the Kondo temperature, $T_K$.
The width of the resonance is proportional to $T_K$ at low
temperatures, growing with $T$ for $T > T_K$.
Its shape is governed by a single interference parameter $q$,
much in the same way as in the noninteracting case.~\cite{Fano61}
The value of $q$ and its spatial variation depends quite sensitively
on details of the underlying surface Green function and the
tunneling matrix elements, which limits the direct applicability
of our results to the experimental data. Nevertheless, we are able
to make some qualitative statements regarding the experiment. In
particular, the indirect interference with the magnetic adatom,
i.e., that due to the tunneling between the STM tip and the
underlying conduction electrons, is found to have a characteristic
range of the order of two lattice spacings, consistent with the
limited spatial extent of the Kondo resonance in the experiment.
In addition, the spatial dependence of the differential
conductance as measured for Co on Au(111)~\cite{STM98_co}
is shown to be nongeneric, indicating a crucial dependence on
the microscopic details of the (111) surface of gold.
Finally, since the widths of the resonant features
in Refs.~\onlinecite{STM98_ce} and \onlinecite{STM98_co}
are both notably larger than the temperature, we conclude
that $T < T_K$ in both experiments.

The remainder of the paper is organized as follows: In
Sec.~\ref{sec:Model} the basic model is introduced, and a general
expression is derived for the differential conductance in the
limit of a weak tunneling current. Sec.~\ref{sec:GF's} then
details our calculation scheme for the dressed impurity Green
function and the impurity-free surface Green function. Simplified
expressions for the first few nonlocal surface Green functions
of our model are provided in Appendix~\ref{appendix:surface_GF}.
Results for the voltage, temperature, and spatial dependence
of the low-voltage differential conductance are presented in
turn in sections~\ref{sec:Point_tunneling} and
\ref{sec:nonlocal_tunneling}, with Sec.~\ref{sec:Point_tunneling}
dedicated to the case of point tunneling between the STM tip
and the underlying substrate conduction electrons, and
Sec.~\ref{sec:nonlocal_tunneling} devoted to the case of a
spatially extended matrix element for tunneling. The main
results of the paper are finally summarized in
Sec.~\ref{sec:Discussion}, followed by a brief comparison
with previous work in Appendix~\ref{appendix:comparison},
and a generalization of our approach to the case of multiple
magnetic adatoms in Appendix~\ref{appendix:multiple_adatoms}.

\section{Model and tunneling current}
\label{sec:Model}

\begin{figure}
\centerline{
\vbox{\epsfxsize=70mm \epsfbox {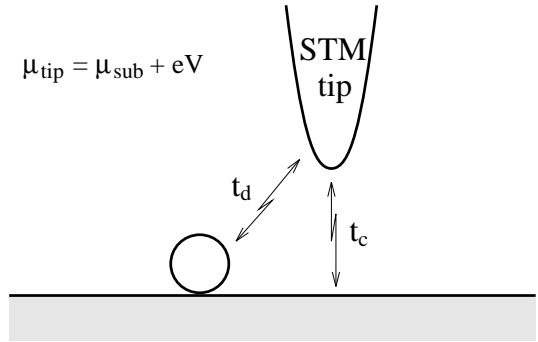}}
}\vspace{10pt}
\caption{Schematic description of the physical system. A metallic surface
with an individual magnetic adatom is probed by an STM tip. The tip couples
separately to the atomic $d$ electrons and to the underlying conduction
electrons via the tunneling matrix elements $t_d$ and $t_c$, respectively.
A voltage bias is applied between the sample and the STM tip, causing a
weak electrical current to flow between the substrate and the tip.}
\label{fig:fig1}
\end{figure}

The system under consideration is shown schematically in
Fig.~\ref{fig:fig1}.
It consists of an individual magnetic adatom, modeled by a $d$ level
with energy $\epsilon_d$ and an on-site repulsion $U$, deposited on
top of a metallic surface. The $d$ electrons hybridize locally with
the underlying conduction electrons via the matrix element $V_{h}$. This
setting is probed by an STM tip which is placed directly above the
surface point $\vec{R}_s$, and which couples separately to the $d$
electrons and to the local conduction electrons at site $\vec{R}_s$
through the tunneling
matrix elements $t_d$ and $t_c$, respectively. The adatom is taken to
be positioned at $\vec{R}_i$. An applied voltage bias between the substrate
and the tip offsets the two chemical potentials, $\mu_{tip} - \mu_{sub} =
eV$, causing a weak electrical current to flow between the substrate
and the tip.

Setting the substrate chemical potential as our reference energy, the
corresponding Hamiltonian takes the form ${\cal H} = {\cal H}_{sub} +
{\cal H}_{tip} + {\cal H}_{tun}$, where
\begin{eqnarray}
{\cal H}_{sub} &=& \sum_{\vec{k}\sigma} \epsilon_{\vec{k}}
           c^{\dagger}_{\vec{k}\sigma}c_{\vec{k}\sigma}
           + \epsilon_d\sum_{\sigma}d^{\dagger}_{\sigma}d_{\sigma}
           + U d^{\dagger}_{\uparrow}d_{\uparrow}
           d^{\dagger}_{\downarrow}d_{\downarrow}\nonumber\\
&+& V_{h} \sum_{\sigma}
           \left \{ d^{\dagger}_{\sigma} \psi_{\sigma}(\vec{R}_i) +
           \psi^{\dagger}_{\sigma}(\vec{R}_i)d_{\sigma} \right \} ,
\label{H_sub}\\
{\cal H}_{tip} &=& \sum_{\vec{k}\sigma} (E_{\vec{k}} + eV)
           a^{\dagger}_{\vec{k}\sigma}a_{\vec{k}\sigma},
\label{H_tip}\\
{\cal H}_{tun} &=& 
           t_c\sum_{\sigma} \left \{ 
           \psi^{\dagger}_{\sigma}(\vec{R}_s) A_{\sigma} +
           A^{\dagger}_{\sigma}\psi_{\sigma}(\vec{R}_s) \right \}
\nonumber\\
&+& t_d \sum_{\sigma} \left \{ d^{\dagger}_{\sigma}A_{\sigma} +
           A^{\dagger}_{\sigma}d_{\sigma} \right \} .
\label{H_tun}
\end{eqnarray}
Here $c^{\dagger}_{\vec{k}\sigma}$ ($a^{\dagger}_{\vec{k}\sigma}$)
creates a bulk (tip) conduction electron with wave number $\vec{k}$
and spin projection $\sigma$, and $d^{\dagger}_{\sigma}$ creates an
atomic $d$ electron with spin $\sigma$. The fermion operators
$\psi_{\sigma}(\vec{r})$ and $A_{\sigma}$ in
Eqs.~(\ref{H_sub})--(\ref{H_tun}) are the local conduction electrons
at point $\vec{r}$ on the surface and at the edge of the STM tip,
respectively. Explicitly,
\begin{eqnarray}
\psi_{\sigma}(\vec{r}) = \sum_{\vec{k}}
     \varphi_{\vec{k}}(\vec{r}) c_{\vec{k}\sigma} ,\\
A_{\sigma} = \sum_{\vec{k}}
     \chi_{\vec{k}} a_{\vec{k}\sigma} ,
\end{eqnarray}
where $\varphi_{\vec{k}}(\vec{r})$ and $\chi_{\vec{k}}$ are the
corresponding bulk and tip single-particle wave functions, evaluated at
point $\vec{r}$ on the surface and at the edge of the tip, respectively.
The atomic energies $U + \epsilon_d$ and $-\epsilon_d$ are assumed to
be positive and large, such that a local moment forms on the adatom.

For zero tunneling, $t_c, t_d = 0$, the local $d$ moment undergoes
Kondo screening below a characteristic temperature $T_K \propto \exp
\left[ -1/\rho_0 J\right]$, where $\rho_0$ is the local surface density
of states at the Fermi level, and $J = 2 V^2_{h} \left [1/|\epsilon_d|
+ 1/(U + \epsilon_d) \right]$ is the effective exchange coupling between
the $d$ moment and the underlying $c^{\dagger}_{\vec{k}\sigma}$
conduction electrons. For nonzero tunneling and a finite voltage
bias, one is dealing in principle with a
nonequilibrium state.~\cite{NEQ_K,NEQ_A} However for a weak
tunneling current, as is the case in the experiments of
Ref.~\onlinecite{STM98_ce} and Ref.~\onlinecite{STM98_co},
the differential conductance essentially probes the zero-tunneling
(equilibrium) Kondo resonance. To see this we note that, for weak
tunneling, it is sufficient to evaluate the tunneling current from the
substrate to the STM tip to second order in $t_c$ and $t_d$. Using
standard diagrammatic techniques one obtains
\begin{equation}
I(V) = \frac{2 e}{\pi \hbar}\sum_{\sigma} \int_{-\infty}^{\infty}\!\!
       \rho_{A\sigma}(\epsilon - eV) \rho_{f\sigma}(\epsilon)
       \left[ f(\epsilon - eV)\!-\!f(\epsilon) \right] d\epsilon ,
\label{current}
\end{equation}
where $f(\epsilon)$ is the Fermi-Dirac distribution function, and
$\rho_{A\sigma}(\epsilon) = -{\rm Im} G_{A\sigma}(\epsilon + i\eta)$
and $\rho_{f\sigma}(\epsilon) = -{\rm Im} G_{f\sigma}(\epsilon + i\eta)$ are
the zero-tunneling spectral functions corresponding to $A_{\sigma}$ and 
\begin{equation}
f_{\sigma} = t_c \psi_{\sigma}(\vec{R}_s) + t_d d_{\sigma} ,
\end{equation}
respectively. Here we have introduced the notation by which the arguments
of $G_{A\sigma}$ and $G_{f\sigma}$ (and thus those of $\rho_{A\sigma}$
and $\rho_{f\sigma}$) are measured relative to their respective chemical
potentials, i.e., $\mu_{tip} = eV$ and $\mu_{sub} = 0$.

All information of the Kondo effect in Eq.~(\ref{current}) is contained
within the retarded $G_{f\sigma}$ propagator,
\begin{eqnarray}
&& \;\;\;\;
G_{f\sigma}(\epsilon + i\eta) =
      t_c^2 G_{\vec{R}_s, \vec{R}_s}(\epsilon + i\eta) +
      G_{\sigma}^{d}(\epsilon + i\eta) \times
\nonumber \\
&&
      \left[t_d + t_c V_{h} G_{\vec{R}_s, \vec{R}_i}
                         (\epsilon + i\eta)\right]\!
      \left[t_d + t_c V_{h} G_{\vec{R}_i, \vec{R}_s}
                         (\epsilon + i\eta) \right] ,
\label{G^f}
\end{eqnarray}
which features the fully dressed impurity ($d$-electron) Green function in
the absence of tunneling, $G_{\sigma}^{d}(\epsilon + i\eta)$, and
the impurity-free surface Green function,
\begin{equation}
G_{\vec{r}_1, \vec{r}_2}(\epsilon + i\eta) = \sum_{\vec{k}}
         \frac{\varphi_{\vec{k}}(\vec{r}_1)
               \varphi^{\ast}_{\vec{k}}(\vec{r}_2)}
         {\epsilon - \epsilon_{\vec{k}} + i\eta} .
\label{G_surface}
\end{equation}
The differential conductance is given in turn by the derivative
of Eq.~(\ref{current}) with respect to $V$. Assuming
$\rho_{A\sigma}(\epsilon) = \rho_A$ is essentially energy
independent on the scale of the voltage bias and the
temperature, one obtains
\begin{equation}
G(V) = -\frac{2 e^2}{\pi \hbar}\rho_A\int_{-\infty}^{\infty}
       \sum_{\sigma}\rho_{f\sigma}(\epsilon + eV)
       \frac{\partial f(\epsilon)}{\partial \epsilon} d\epsilon .
\label{differential_cond}
\end{equation}

Equations~(\ref{G^f})--(\ref{differential_cond}) are a generalization
of Fano's original analysis~\cite{Fano61} to the interacting case. They
express the differential conductance in terms of the fully dressed
impurity Green function, the impurity-free surface Green function, and
the microscopic tunneling parameters of the model. The remainder of the
paper is devoted to evaluation and analysis of these equations, starting
with the Green functions $G^d_{\sigma}(\epsilon + i\eta)$ and
$G_{\vec{r}_1, \vec{r}_2}(\epsilon + i\eta)$.

\section{Impurity and surface Green functions}
\label{sec:GF's}
\subsection{Impurity Green function}
We have evaluated the impurity Green function in the limit $U \to \infty$,
where double occupancy is forbidden on the magnetic adatom. Calculations
were performed using the non-crossing approximation~\cite{Bickers87} (NCA),
which is a self-consistent perturbation theory about the atomic limit.
This approach is known to provide a good quantitative description of the
temperature range $T \agt T_K$, and has the advantage of being capable
of treating systems with realistically small Kondo temperatures. This is
to be contrasted with Quantum Monte Carlo approaches, which can not go
to very low temperatures. As a large-$N$ theory, however, the NCA fails
to recover Fermi-liquid behavior at low temperatures,~\cite{Bickers87}
and it overshoots the unitary limit for $T \alt T_K$ in the $N = 2$,
nondegenerate case. To avoid these pathologies of the NCA, we
restrict attention in this paper to the range $T \geq T_K$.

\subsection{Surface Green function}
Due to the local nature of the hybridization between the $d$ and
conduction electrons in Eq.~(\ref{H_sub}), only the local
conduction-electron density of states is relevant to the
impurity Green function, $G^d_{\sigma}(\epsilon + i\eta)$.
Conventionally, this allows one to parameterize the band by a single
function for the density of states, often chosen for convenience to
have a Lorentzian, box, or Gaussian form. In contrast, calculation of
the differential conductance for $\vec{R}_s \neq \vec{R}_i$ requires
detailed information of the band dispersion. In particular,
a consistent theory for the energy and spatial dependence
of $G(V)$ requires one to start from a microscopic description of
the underlying conduction band.

As a generic model for the substrate conduction band, we consider a
simple-cubic tight-binding Hamiltonian, with open boundary conditions at
the surface. Setting the lattice spacing as our unit length, the lattice
is described by the integer grid $\vec{r}_i = (x_i, y_i, z_i)$ with
$z_i > 0$, such
that the first (surface) monolayer corresponds to $z_i = 1$. In each of the
$x$ and $y$ direction we take periodic boundary conditions, i.e., $x_i$ and
$y_i$ are equivalent to $x_i + N$ and $y_i + N$, respectively. In the $z$
direction we apply open boundary conditions, namely, $c_{\vec{r}_i\sigma}$
(the conduction-electron annihilation operator at site $\vec{r}_i$)
vanishes for $z_i = 0$ and $z_i = N$. The corresponding Hamiltonian reads
\begin{equation}
{\cal H}_{band} = t\!\sum_{<i,j>, \sigma}\!\left \{
         c^{\dagger}_{\vec{r}_i\sigma} c_{\vec{r}_j\sigma} +
         c^{\dagger}_{\vec{r}_j\sigma} c_{\vec{r}_i\sigma} \right \} ,
\label{H_tight-binding}
\end{equation}
where $<\!i,j\!>$ denotes nearest-neighbor lattice sites. For
$\mu_{sub} = 0$, the case considered here, the Hamiltonian of
Eq.~(\ref{H_tight-binding}) describes a half-filled band.

The tight-binding Hamiltonian of Eq.~(\ref{H_tight-binding}) is
diagonalized by converting to the single-particle basis
\begin{equation}
c_{\vec{k}\sigma} = \sqrt{\frac{2}{N^{3}}}\sum_{i}
         c_{\vec{r}_i\sigma} e^{-i(k_x x_i + k_y y_i)}\sin(k_z z_i) .
\end{equation}
Here $\vec{k}$ takes the values $(2n_x, 2n_y, n_z) \pi/N$, with the
integers $0 \le n_x, n_y < N$ and $1 \le n_z < N$. The corresponding
single-particle energies take the form
\begin{equation}
\epsilon_{\vec{k}} = 2t\!\sum_{l = x,y,z}\!\cos(k_l) .
\end{equation}

Since the tight-binding wave functions $\varphi_{\vec{k}}(\vec{r})$ are
confined to a discrete set of lattice sites, the Green function of
Eq.~(\ref{G_surface}) is meaningful only in the case where both
$\vec{r}_1$ and $\vec{r}_2$ are lattice sites. Taking the
thermodynamic limit, $N \to \infty$, one thus obtains
\begin{eqnarray}
&&\;\;\;\;\;\;\;
         G_{\vec{r}_1,\vec{r}_2}(\epsilon + i\eta) = \frac{2}{\pi^3}
         \int_{0}^{\pi}\!\!dk_x\int_0^{\pi}\!\!dk_y\!
         \int_{0}^{\pi}\!\!dk_z
\label{explicit_bulk_GF} \\
&&\times \cos[k_x (x_1\!-\!x_2)] \cos[k_y (y_1\!-\!y_2)]
         \frac{\sin(k_z z_1)\sin(k_z z_2)}
         {\epsilon - \epsilon_{\vec{k}} + i\eta} ,
\nonumber
\end{eqnarray}
from which the surface Green function follows by setting $z_1 = z_2 = 1$:
\begin{eqnarray}
&&G_{lm}(\epsilon + i\eta) = \frac{2}{\pi^3}
         \int_{0}^{\pi}\!\!dk_x\int_0^{\pi}\!\!dk_y\!
         \int_{0}^{\pi}\!\!dk_z\nonumber\\
&&\;\;\;\;\;
         \times
         \cos(k_x l)\cos(k_y m) \frac{\sin^2(k_z)}
         {\epsilon - \epsilon_{\vec{k}} + i\eta} .
\label{explicit_surface_GF}
\end{eqnarray}
Here and in the remainder of the paper we label the surface Green
function by two integers: $l = x_1 - x_2$ and $m = y_1 - y_2$. We further
note that $G_{lm}(\epsilon + i\eta)$ and $G_{ml}(\epsilon + i\eta)$
are identical by symmetry, and are only dependent upon the distances
$|l|$ and $|m|$.

\begin{figure}
\centerline{
\vbox{\epsfxsize=75mm \epsfbox {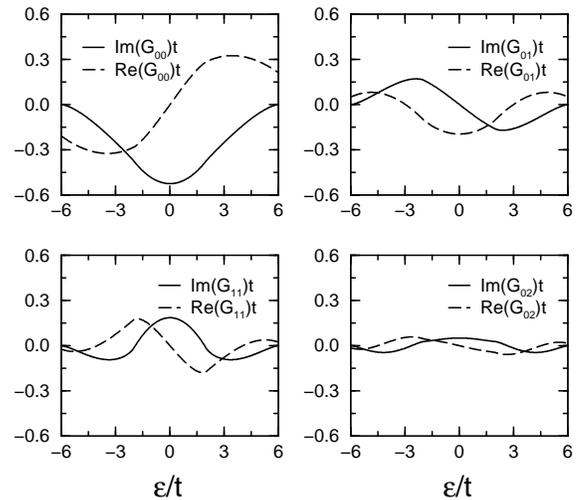}}
}\vspace{5pt}
\caption{Real and imaginary parts of the tight-binding surface Green
function $G_{lm}(\epsilon + i\eta)$, for different values of $(l,m)$.}
\label{fig:fig2}
\end{figure}

Equation~(\ref{explicit_surface_GF}) can be further simplified to just a
single integration by exploiting the relation to the local tight-binding
Green function for a two-dimensional, simple-cubic lattice:~\cite{2DDOS}
\begin{equation}
G_{2D}(\zeta) = \frac{2}{\pi \zeta} K[ (4t/\zeta)^2 ] .
\end{equation}
Here $K(\zeta)$ is the complete elliptic integral of the first
kind, analytically continued to the upper and lower half
planes.~\cite{Abramowitz} Specifically, the local surface
Green function, $G_{00}$, is conveniently expressed as
\begin{equation}
G_{00}(\epsilon + i\eta) = \frac{2}{\pi}\int_{0}^{\pi}\!dk \sin^2(k)
          G_{2D}[\epsilon - 2t\cos(k) + i\eta] ,
\label{concise_G_00}
\end{equation}
with analogous expressions for the first few nonlocal Green functions
(see Appendix~\ref{appendix:surface_GF}).

Figure~\ref{fig:fig2} displays the first four surface Green functions.
Both the real and imaginary parts of $G_{lm}(\epsilon + i\eta)$ decay
as a function of $\sqrt{l^2 + m^2}$, alternating between even and odd
functions of energy. For even (odd) $|l| + |m|$, the real part of
$G_{lm}(\epsilon + i\eta)$ is odd (even) in energy, while the imaginary
part is even (odd). As a result, $G_{lm}(0 + i\eta)$ is purely
imaginary when $|l| + |m|$ is even, and purely
real when $|l| + |m|$ is odd. This has a profound effect on the spatial
dependence of the low-voltage differential conductance for this
model, as a purely real
$G_{lm}(0 + i\eta)$ gives rise to resonant enhancement of the differential
conductance for $\vec{R}_s - \vec{R}_i = (l, m, 0)$. By contrast,
an imaginary $G_{lm}(0 + i\eta)$ can result both in an antiresonance
or an asymmetric Fano resonance, depending on details of the
tunneling matrix elements
$t_c$ and $t_d$. We shall return to this point in great detail below.
Finally, we note that the imaginary part of
$G_{lm}(0 + i\eta)$ is nonzero only in the range $-6t < \epsilon < 6t$,
and that the van-Hove singularities at $\epsilon = \pm 2t$ are considerably
smoother at the surface than deep inside the bulk.

\begin{figure}
\centerline{
\vbox{\epsfxsize=75mm \epsfbox {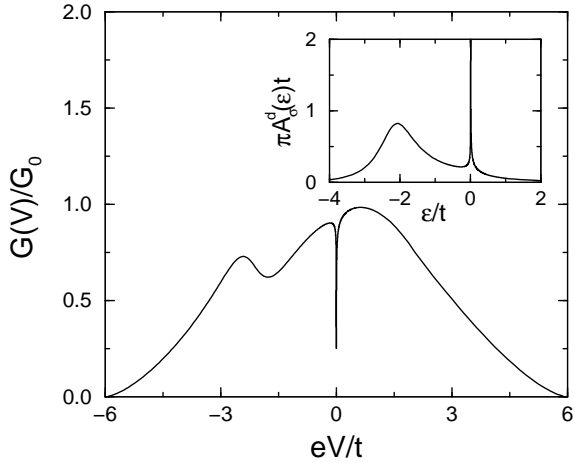}}
}\vspace{5pt}
\caption{The differential conductance, $G(V)$, as a function of voltage,
for an STM tip placed directly above the magnetic adatom. Here $T = T_K$,
$t_d = 0$, and $G_0 = 4e^2 t_c^2 \rho_0 \rho_A/\hbar$. The impurity model
parameters are $\epsilon_d/t = -1.67$, $\Gamma/t = 0.4$, and $U = \infty$,
corresponding to a Kondo temperature of $T_K/t = 10^{-3}$. The impurity
Green function, whose spectral part $A_{\sigma}^d(\epsilon) =
-(1/\pi) {\rm Im} \left \{ G_{\sigma}^{d}(\epsilon + i\eta) \right\}$
is plotted in the inset, was computed using the NCA. In addition to the
standard broad feature near the $d$ level, a sharp antiresonance is
seen in $G(V)$ at zero bias, corresponding to the Abrikosov-Suhl
resonance in the impurity spectral function.}
\label{fig:fig3}
\end{figure}

\section{Results for point tunneling}
\label{sec:Point_tunneling}
We have evaluated the differential conductance according to
Eqs.~(\ref{G^f})--(\ref{differential_cond}), within the tight-binding
model of Eq.~(\ref{H_tight-binding}) for the underlying conduction band.
$\psi^{\dagger}_{\sigma}(\vec{R}_i)$ was identified in this picture with
the creation of an electron at the lattice site $\vec{R}_i = (0,0,1)$.
The local density of states used in computing the impurity Green
function was taken accordingly to be $\rho(\epsilon) =
-\frac{1}{\pi}{\rm Im} \left \{ G_{00}(\epsilon + i\eta) \right\}$,
where $G_{00}$ is given by Eq.~(\ref{concise_G_00}). The
corresponding density of states at the Fermi level is equal to
$\rho_0 = \rho(0) = 0.525/t$.
Focusing on the case where the STM tip is positioned directly above the
lattice site $\vec{R}_s = (l,m,1)$, each of the surface Green functions
$G_{\vec{R}_s,\vec{R}_i}(\epsilon + i\eta)$ and
$G_{\vec{R}_i,\vec{R}_s}(\epsilon + i\eta)$ in Eq.~(\ref{G^f}) was
identified with $G_{lm}(\epsilon + i\eta)$.

\subsection{Tip placed above the magnetic adatom}
We begin with an STM tip positioned directly above the magnetic
adatom. As the temperature is lowered down to $T_K$, a sharp
resonant structure develops in the differential conductance
at zero bias, in addition to the standard broad feature near
the $d$ level. This is demonstrated Fig.~\ref{fig:fig3} for
$T = T_K$ and $t_d = 0$. The sharp antiresonance seen at zero
bias in this case directly corresponds to the Abrikosov-Suhl
resonance in the impurity spectral function, which is plotted
for comparison in the inset of Fig.~\ref{fig:fig3}.
Here and throughout the paper, the
impurity Green function was calculated within the NCA, using
the impurity model parameters $\epsilon_d/t = -1.67$,
$\Gamma/t \equiv \pi\rho_0 V_h^2/t = 0.4$, and
$U = \infty$.~\cite{comment_on_unitary_limit}
The corresponding Kondo temperature, $T_K/t = 10^{-3}$, was defined
as the temperature at which the ``resistivity''
integral,~\cite{comment_on_resistivity}
\begin{equation}
R(T) = \left [ \int_{-\infty}^{\infty}
       \frac{1}{ {\rm Im}\!\left\{ G_{\sigma}^{d}(\epsilon + i\eta)\right\} }
       \frac{\partial f(\epsilon)}{\partial\epsilon}d\epsilon\right ]^{-1} ,
\end{equation}
reduces to 50\% of its zero-temperature limit. This definition of $T_K$
agrees to within a factor of order unity with the half-width of
the Abrikosov-Suhl resonance in the impurity spectral function.

\begin{figure}
\centerline{
\vbox{\epsfxsize=75mm \epsfbox {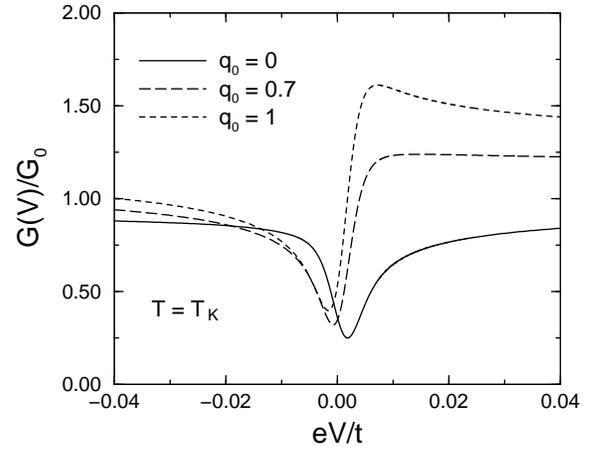}}
}\vspace{5pt}
\caption{The low-voltage differential conductance for an STM tip
placed directly above the magnetic adatom. Here $T = T_K$ and
$q_0 = t_d/\pi \rho_0 t_c V_h$, with all other model parameters
as in Fig.~\ref{fig:fig3}. With increasing $|q_0|$, $G(V)$
evolves from an antiresonance for $|q_0| \ll 1$, to an asymmetric
Fano resonance for intermediate $|q_0|$, to a full resonance
for $|q_0| \gg 1$ (not shown).}
\label{fig:fig4}
\end{figure}

Focusing hereafter on the low-voltage resonant structure,
Fig.~\ref{fig:fig4} depicts the resonance's dependence
on $q_0 = t_d/\pi \rho_0 t_c V_h$, which plays the role of
Fano's interference parameter $q$ in this case. Here
$G_0 = 4e^2 t_c^2 \rho_0 \rho_A/\hbar$ is the
zero-temperature conductance in the absence of an impurity.
With increasing $|q_0|$, the low-voltage differential conductance
evolves in Fig.~\ref{fig:fig4} from an antiresonance for $|q_0| \ll 1$,
to an asymmetric Fano resonance for intermediate $|q_0|$, to a full
resonance for $|q_0| \gg 1$, much in the same way as in the
noninteracting case. The
antiresonance for $q_0 = 0$ resembles that for a Ce adatom on
Ag(111),~\cite{STM98_ce} supporting the interpretation of
Li {\em et al.} that the observed lineshape stems from
Kondo screening of the Ce moment with only weak direct coupling
between the tip and the adatom. The $q_0 = 1$ curve is similar in
turn to the Fano resonance observed by Madhavan {\em et al.} for a
Co adatom on Au(111),~\cite{STM98_co} while larger values of $q_0$
display too shallow a dip and too high a peak as compared to the
data of Ref.~\onlinecite{STM98_co}.

\begin{figure}
\centerline{
\vbox{\epsfxsize=75mm \epsfbox {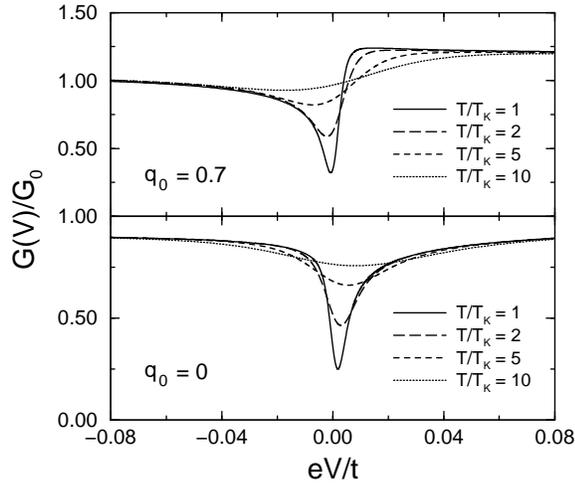}}
}\vspace{5pt}
\caption{Temperature dependence of $G(V)$, for an STM tip placed directly
above the magnetic adatom. Here $q_0 = 0$ and $q_0 = 0.7$ in the lower and
upper graphs, respectively. All other model parameters as in
Fig.~\ref{fig:fig3}.}
\label{fig:fig5}
\end{figure}

Figure~\ref{fig:fig5} shows the temperature dependence of $G(V)$,
for the two representative values of $q_0 = 0$ and $q_0 = 0.7$. The
main effect of a temperature is to broaden and smear the resonant
structure in $G(V)$, whose width grows according to
$T$ for $T > T_K$. Notice the significant reduction in the
overall resonance height by the time $T/T_K \sim 10$.
This strong temperature dependence of the differential conductance has
two contributions: (i) A rapid decrease in the Abrikosov-Suhl resonance
in the impurity spectral function
with increasing temperature, and (ii) Further smearing of the
impurity Green function due to the convolution with the derivative
of the Fermi-Dirac function in Eq.~(\ref{differential_cond}). The
fact that the well-developed features in Refs.~\onlinecite{STM98_ce}
and \onlinecite{STM98_co} have widths that are notably larger than
$T$ is thus a clear indication that $T < T_K$ in both experiments.

\subsection{Spatial dependence of the differential conductance}

\begin{figure}
\centerline{
\vbox{\epsfxsize=70mm \epsfbox {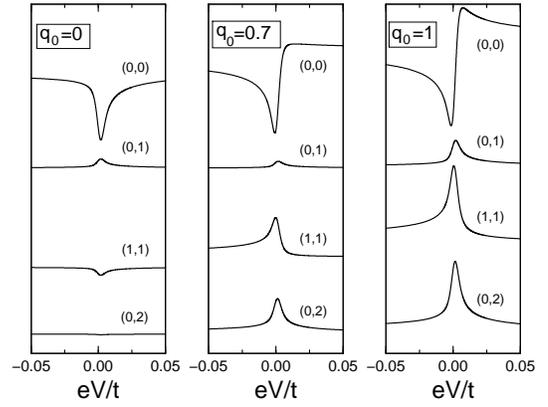}}
}\vspace{5pt}
\caption{Spatial dependence of the low-voltage
differential conductance, $G(V)$, for $T = T_K$
and different values of $q_0 = t_d/\pi \rho_0 t_c V_h$. All impurity model
parameters are as in Fig.~\ref{fig:fig3}. The individual curves are
offset by one unit each, according to the lateral distance from the
impurity site.}
\label{fig:fig6}
\end{figure}

As stated above, the novel aspect of STM spectroscopy of the Kondo
effect is the ability to spatially resolve the electronic structure
around the magnetic adatom. In Fig.~\ref{fig:fig6} we have plotted
the low-voltage differential conductance as a function of $(l, m)$,
for the idealized case where the STM tip is placed directly above
the lattice site $\vec{R}_s = (l, m, 1)$. Here
each set of curves corresponds to a different fixed value of
$q_0 = t_d/\pi \rho_0 t_c V_h$, which no longer corresponds for
$\vec{R}_s \neq \vec{R}_i$ to Fano's interference parameter $q$.
The latter also depends on the real and imaginary parts of
$G_{\vec{R}_s,\vec{R}_i} (0 + i\eta)$, as discussed below.
Physically one should recall, though, that $q_0 \propto t_d$
decays to zero with increasing distance from the magnetic adatom,
hence $G(V)$ always tends to the $q_0 = 0$ curve as the lateral
distance from the adatom is increased.

Fixing the value of $q_0$ for the time being and increasing the lateral
distance from the magnetic adatom, the low-bias differential conductance
approaches the asymptotic curve
\begin{equation}
G(V) = G_0 \left [ 1 + \Gamma q_0^2\!\int_{-\infty}^{\infty}\!\!
       {\rm Im}\!\left\{ G_{\sigma}^{d}(\epsilon + i\eta) \right \}
       \frac{\partial f(\epsilon)}{\partial\epsilon}d\epsilon\right] .
\label{fixed_q_large_r}
\end{equation}
This result stems from the decaying nature of $G_{lm}(0 + i\eta)$,
along with the fact that $G_{00}(\epsilon + i\eta) = -i \pi \rho_0$
is approximately constant on the scale of the voltage bias and the
temperature. As seen in Fig.~\ref{fig:fig6},
Eq.~(\ref{fixed_q_large_r}) is approached at a lateral distance of
about two lattice spacings from the magnetic adatom. For larger distances
the resonance height is basically proportional to the residual
coupling to the magnetic adatom squared. Specifically, there are no
visible traces of the Kondo resonance for $q_0 = 0$ when the tip and the
adatom are two lattice spacings apart [the curve $(l, m) = (0, 2)$ in
Fig.~\ref{fig:fig6}]. Such a limited spatial extent of the Kondo
resonance in $G(V)$ is consistent with the one seen experimentally for
Co on Au(111),~\cite{STM98_co} indicating that $q_0$ is effectively
zero in the experiment above a lateral distance of about 10\AA.
Indeed, the low-bias resonant structure is also mostly gone by
lateral distance of 10\AA\ for Ce on Ag(111),~\cite{STM98_ce}
even though the Ag(111) surface state at $-70$meV does not
fully set in before a distance of about 40\AA.

The resonant enhancement of $G(V)$ at a lateral distance of one
lattice spacing, which occurs for any value of $q_0$, is not seen
in the experiment.
This feature of the calculated differential conductance
is traced back to the fact that $G_{01}(0 + i\eta)$ is purely
real for our tight-binding model, resulting in a differential
conductance $G(V)$ that is once again given by
Eq.~(\ref{fixed_q_large_r}), but with
\begin{equation}
q_0 \to q_0 + \frac{1}{\pi \rho_0}{\rm Re} \left\{G_{01}(0 + i\eta)\right\} 
      = q_0 - 0.372
\end{equation}
[see Eqs.~(\ref{G^f}) and (\ref{differential_cond})]. Thus, irrespective
of the actual $q_0$ that applies to a lateral distance of one lattice
spacing in the experiment of Madhavan {\em et al.}, our present model
fails to recover the dip-like structure seen experimentally at such
a distance for Co on Au(111).~\cite{STM98_co}

That $G_{01}(0 + i\eta)$ is purely real is a generic feature of half-filled
nearest-neighbor tight-binding models on bipartite lattices. It is lost,
however, when the system is away from half filling, or upon inclusion
of a next-nearest-neighbor hopping term. This suggests that the dip-like
structure seen experimentally up to a distance of $\sim$10\AA\ from
the Co adatom~\cite{STM98_co} is due to a qualitative difference in
the underlying surface Green function for Au(111). Indeed, the (111)
surface of gold is known to have a herringbone reconstruction with
regions of fcc and hcp ordering,~\cite{herringbone} which differs
from the simple-cubic structure considered here. This may support
the interpretation of
a qualitative difference in the underlying surface Green function.

It should be noted, though, that such an interpretation for Au(111)
demands that ${\rm Im} \{ G_{\vec{R}_i, \vec{R}_s}(0 + i\eta) \}$
is never small compared to
$\pi \rho_0 q_0 + {\rm Re} \{ G_{\vec{R}_i, \vec{R}_s}(0 + i\eta) \}$
in the relevant range in $\vec{R}_s - \vec{R}_i$. In particular,
this rules out any oscillatory behavior of
${\rm Im} \{ G_{\vec{R}_i, \vec{R}_s}(0 + i\eta) \}$
as a function of $\vec{R}_s - \vec{R}_i$ in this range. A brief
examination of Fig.~\ref{fig:fig2} reveals that this condition is
actually quite restrictive, at least for the tight-binding model of
Eq.~(\ref{H_tight-binding}): Only in a narrow window of band fillings
is ${\rm Im} \{ G_{lm}(0 + i\eta) \}$ consistently negative for all
$|l| + |m| \leq 2$. We thus conclude that the spatial dependence
measured by Madhavan {\em et al.} is certainly not generic, but
depends on details of the underlying band.

\section{Finite spatial extent of the tunneling matrix element}
\label{sec:nonlocal_tunneling}

Thus far we have considered an idealized point tunneling between the STM
tip and the substrate conduction electrons at point $\vec{R}_s$. In practice,
however, the tunneling matrix element has a finite spatial extent about
$\vec{R}_s$, which is reflected in the different lineshapes that
are observed when the tip is removed from the Co adatom in opposite
directions.~\cite{STM98_co} On the level of the model, a spatially
extended tunneling matrix element is accounted for by replacing
$\psi_{\sigma}(\vec{R}_s)$ in the first term of Eq.~(\ref{H_tun}) with a
weighed sum over the conduction-electron degrees of freedom around
$\vec{R}_s$:
\begin{equation}
\psi_{\sigma}(\vec{R}_s) \to
\sum_{\vec{r}} w_{\vec{r}}\psi_{\sigma}(\vec{R}_s + \vec{r}) .
\label{superposition}
\end{equation}
Here we use the convention
\begin{equation}
\sum_{\vec{r}} |w_{\vec{r}}|^2 = 1 ,
\label{w_normalization}
\end{equation}
which fixes the separation of the local tunneling matrix element at
each $\vec{r}$ into $t_{\vec{r}} = t_c w_{\vec{r}}$. For the
tight-binding model of Eq.~(\ref{H_tight-binding}), the sum over $\vec{r}$
in Eq.~(\ref{superposition}) extends over those lattice sites close to
the STM tip.

\begin{figure}
\centerline{
\vbox{\epsfxsize=65mm \epsfbox {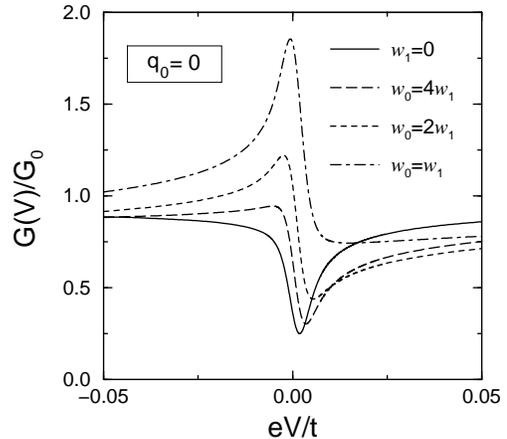}}
}\vspace{5pt}
\caption{The differential conductance, $G(V)/G_0$, as a function of
$w_1/w_0$, for $T = T_K$, $q_0 = 0$, and an STM tip placed directly above
the magnetic adatom. Here both $w_0$ and $w_1$ are assumed to be real
and positive. Note that $G_0$ itself varies as a function of $w_1/w_0$
[see Eq.~(\ref{G_0_general})], taking the values
$G_0 \hbar/4e^2 t_c^2 \rho_0 \rho_A = 1, 0.838, 0.598$, and $0.356$
for $w_1/w_0 = 0, 0.25, 0.5$, and $1$, respectively. All impurity model
parameters are as in Fig.~\ref{fig:fig3}. As the ratio $w_1/w_0$
is increased from zero to one, $G(V)$ evolves from an antiresonance to a
resonance, corresponding to an increase in the effective interference
parameter $q$ [see Eq.~(\ref{q_eff}) and accompanying text].}
\label{fig:fig7}
\end{figure}

Upon substituting Eq.~(\ref{superposition}) into the tunneling
Hamiltonian of Eq.~(\ref{H_tun}), the differential conductance
remains given by Eqs.~(\ref{G^f})--(\ref{differential_cond}),
but with the following modifications to Eq.~(\ref{G^f}):
\begin{mathletters}
\begin{eqnarray}
G_{\vec{R}_s,\vec{R}_s} &\to&
       \sum_{\vec{r}, \vec{r}'} w_{\vec{r}} w^{\ast}_{\vec{r}'}
       G_{\vec{R}_s + \vec{r},\vec{R}_s + \vec{r}'} ,\\
G_{\vec{R}_s,\vec{R}_i} &\to& \sum_{\vec{r}} w_{\vec{r}}
       G_{\vec{R}_s + \vec{r},\vec{R}_i} ,\\
G_{\vec{R}_i,\vec{R}_s} &\to& \sum_{\vec{r}}
       w^{\ast}_{\vec{r}} G_{\vec{R}_i, \vec{R}_s + \vec{r}} .
\end{eqnarray}
\label{sub_G's}
\end{mathletters}

\noindent Accordingly, the zero-temperature conductance in the absence
of an adatom is equal to
\begin{equation}
G_0 = -\frac{4e^2 t_c^2}{\pi \hbar} \rho_A {\rm Im} \left \{
       \sum_{\vec{r}, \vec{r}'} w_{\vec{r}} w^{\ast}_{\vec{r}'}
       G_{\vec{R}_s + \vec{r},\vec{R}_s + \vec{r}'}(0 + i\eta) \right \} ,
\label{G_0_general}
\end{equation}
which properly reduces to $G_0 = 4e^2 t_c^2\rho_0\rho_A/\hbar$
in the limit of point tunneling.

To examine the effect of a finite spatial extent in the tunneling matrix
element, we go back to the tight-binding model of
Eq.~(\ref{H_tight-binding}), and to the case where the tip is placed
directly above the magnetic adatom, i.e., $\vec{R}_s = (0,0,1)$. In
addition to tunneling between the tip and the lattice point $\vec{R}_s$,
we introduce a nonzero tunneling matrix element to each of the four
surface nearest-neighbors of $\vec{R}_s$. Restricting attention to the
isotropic case, one is left with two different $w_{\vec{r}}$ parameters:
$w_0$ for the tunneling matrix element to the lattice point
$\vec{R}_s$, and $w_1$ for the tunneling matrix element to each
of its four surface nearest neighbors. The normalization condition,
Eq.~(\ref{w_normalization}), then reads
\begin{equation}
|w_1| = \frac{1}{2} \sqrt{1 - |w_0|^2} .
\end{equation}

\begin{figure}
\centerline{
\vbox{\epsfxsize=65mm \epsfbox {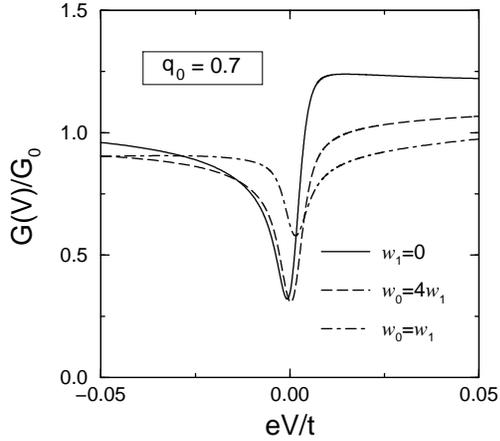}}
}\vspace{5pt}
\caption{Same as Fig.~\ref{fig:fig7}, but for $q_0 = 0.7$.
As the ratio $w_1/w_0$ is increased from zero to one, $G(V)$ evolves
from an asymmetric Fano resonance to an antiresonance, corresponding to
a decrease in the effective interference parameter $q$ [see
Eq.~(\ref{q_eff}) and accompanying text].}
\label{fig:fig8}
\end{figure}

Figures~\ref{fig:fig7} and \ref{fig:fig8} depict the evolution of
the low-voltage differential conductance $G(V)/G_0$ as a function
of $w_1/w_0$, for the
two representative values of $q_0 = 0$ and $q_0 = 0.7$. Here we have
focused for simplicity on the case where $w_0$ and $w_1$ are both real
and positive, yet the qualitative picture does not depend on this
choice. The dramatic effect that a nonzero $w_1$ has on the differential
conductance in this case can be understood within Fano's interference
picture. Substituting Eqs.~(\ref{sub_G's}) into Eq.~(\ref{G^f})
and taking $w_0$ and $w_1$ to be real, the effective interference
parameter {\em a la} Fano is given for $\vec{R}_s = \vec{R}_i$ by
minus the ratio of the real and imaginary parts of
\begin{equation}
t_d + t_c V_h \left [ w_0 G_{00}(0 + i\eta) + 4 w_1 G_{01}(0 + i\eta)
      \right] .
\label{q_ri=rs}
\end{equation}
Here Eq.~(\ref{q_ri=rs}) corresponds to the resulting expression in each
of the square
brackets of Eq.~(\ref{G^f}). Using $G_{00}(0 + i\eta) = -i\pi \rho_0$
and $G_{01}(0 + i\eta) = -0.372 \pi \rho_0$, as is appropriate for the
model of Eq.~(\ref{H_tight-binding}), one obtains
\begin{equation}
q = (q_0 - 1.49w_1)/w_0 .
\label{q_eff}
\end{equation}

For $q_0 = 0$, Eq.~(\ref{q_eff}) reduces to $-1.49(w_1/w_0)$, which
varies from $q = 0$ to $q = -1.49$ in going from $w_1 = 0$
to $w_1 = w_0$. This strong change in $q$ produces the transition
from an antiresonance to a resonance in the differential
conductance of Fig.~\ref{fig:fig7}. Similarly for $q_0 = 0.7$,
Eq.~(\ref{q_eff}) varies from $q = 0.7$ to $q = 0.075$ in
going from $w_0 = 1$ to $w_0 = w_1$, which causes the transition from
an asymmetric Fano resonance to an antiresonance in
Fig.~\ref{fig:fig8}.

\begin{figure}
\centerline{
\vbox{\epsfxsize=65mm \epsfbox {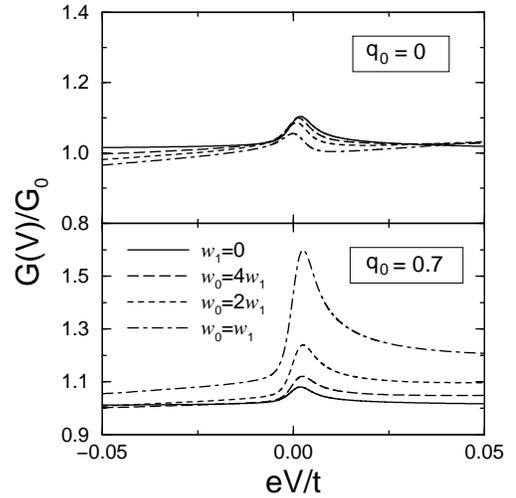}}
}\vspace{5pt}
\caption{The differential conductance, $G(V)/G_0$, as a function of
$w_1/w_0$, for an STM tip one lattice spacing removed from the
magnetic adatom [i.e., $\vec{R}_s = (\pm 1,0,1)$ or $(0,\pm 1, 1)$].
As in Fig.~\ref{fig:fig7}, $w_0$ and $w_1$ are assumed to
be real and positive, with all impurity model parameters the same
as in Fig.~\ref{fig:fig3}. For both $q_0 = 0$ and $q_0 = 0.7$,
the differential conductance continues to feature a resonance for
all $0 \leq w_1/w_0 \leq 1$.}
\label{fig:fig9}
\end{figure}

Repeating the same argumentation for the case where $\vec{R}_s$ and
$\vec{R}_i$ are one lattice spacing apart, i.e., $\vec{R}_s = (\pm 1,0,1)$
or $\vec{R}_s = (0,\pm 1, 1)$, Eq.~(\ref{q_ri=rs}) is modified to
\begin{equation}
t_d + t_c V_h \left [ w_0 G_{01} + w_1 G_{00}
      + 2 w_1 G_{11} + w_1 G_{02} \right] ,
\end{equation}
with all Green functions evaluated at zero frequency. Using
$G_{11}(0 + i\eta) = 0.354 i \pi \rho_0$ and $G_{02}(0 + i\eta) =
0.097 i \pi \rho_0$ one obtains
\begin{equation}
q = ( 5.1 q_0 - 1.9 w_0)/w_1 ,
\label{q_eff_nl}
\end{equation}
which is large in magnitude throughout the range $0 < w_1 \leq w_0$,
for both $q_0 = 0$ and $q_0 = 0.7$. Given the large value of $|q|$,
one expects the differential conductance to continue to show a
resonance for both values of $q_0$ and all $0 \leq w_1/w_0 \leq 1$,
which is precisely what is seen in Fig.~\ref{fig:fig9}.

Thus, while the inclusion of nonlocal tunneling between the
STM tip and the underlying substrate conduction electrons
obviously increases the parametric dependence of $q$,
it does not necessarily assist in producing a dip in the
low-voltage differential conductance when the tip and the
adatom are one lattice spacing apart.

\section{Discussion}
\label{sec:Discussion}

We begin our discussion with the case of an STM tip placed directly above
the magnetic adatom, depicted in Figs.~\ref{fig:fig3}--\ref{fig:fig5} and
Figs.~\ref{fig:fig7}--\ref{fig:fig8}. Similar to the noninteracting case,
the shape of the Kondo resonance in the low-temperature,
low-voltage differential conductance is governed by a single
interference parameter $q$, which depends both on the
ratios of the tunneling matrix elements and on details of the
impurity-free surface Green function. Specifically, in
Fig.~\ref{fig:fig4} $q$ is equal to $q_0 = t_d/\pi \rho_0 t_c V_h$,
whereas in Figs.~\ref{fig:fig7}--\ref{fig:fig8} it is modified
according to Eq.~(\ref{q_eff}). The effect of a temperature is to
rapidly broaden and smear the Kondo resonance in $G(V)$, whose
width grows according to $T$ for $T > T_K$. This behavior stems both
from the standard convolution with the derivative of the Fermi-Dirac
distribution function in Eq.~(\ref{differential_cond}), and from
the rapid decrease in the Abrikosov-Suhl resonance with increasing
temperature. Indeed, the qualitative difference in the low-temperature,
low-voltage differential conductance for a magnetic adatom as compared
to that of a conventional noninteracting resonance is in the strong
energy and temperature dependence of the impurity self-energy, whose
real and imaginary parts cannot be regarded a constant.

Experimentally, the antiresonance observed for Ce on
Ag(111)~\cite{STM98_ce} is similar to the $q_0 = 0$ curve in
Fig.~\ref{fig:fig4}. Assuming point tunneling between the tip
and the underlying conduction electrons, this supports the
interpretation of weak direct tunneling between the tip and
the Ce adatom,~\cite{STM98_ce} which
requires at the same time that $G_{\vec{R}_i\vec{R}_i}(0 + i\eta)$
is mostly imaginary. While plausible, this scenario is certainly
not exclusive, as there are various other ways to obtain
$q \approx 0$ without resorting to a negligible coupling to the
adatom. For example, the curve $w_0 = w_1$ in Fig.~\ref{fig:fig8}
(corresponding to a spatially extended tunneling) is also
characterized by $|q| \ll 1$, although $q_0 = 0.7$ is by no
means small. Likewise, while the similarity between the $q_0 = 1$
curve in Fig.~\ref{fig:fig4} and the Fano resonance for Co on
Au(111)~\cite{STM98_co} is suggestive of comparable contributions
from the tunneling to the adatom and to the underlying conduction
electrons, one cannot rule out other combinations for which
$q_0 \approx 0$. For example, setting $q_0 \approx 0$ and
$w_0/ w_1 \approx -1.5$ in Eq.~(\ref{q_eff}) also results
in $q \approx 1$. On the other hand, the fact that the
well-developed features in Refs.~\onlinecite{STM98_ce}
and \onlinecite{STM98_co} have characteristic widths that
are considerably larger than the temperature is a clear
indication that $T < T_K$ in both experiments.

More detailed information about the underlying electronic
structure is contained in the spatial variation of the
differential conductance, as measured, for example, in
Ref.~\onlinecite{STM98_co} for a Co adatom on Au(111). Here,
although we have considered a particular tight-binding model
for the impurity-free surface, there are some qualitative
statements we can make with regard to the experiment.
Primarily, as seen in Fig.~\ref{fig:fig6}, the indirect interference
with the magnetic adatom, i.e., that due to the tunneling between
the STM tip and the underlying conduction electrons, is suppressed
above a lateral distance of about two lattice spacings from the
adatom. Hence the characteristic range for the indirect interference
with the adatom is of the order of two lattice spacings. While
this range may certainly depend both on details of the
underlying band and on the presence of nonlocal tunneling between
the tip and the substrate conduction electrons, we expect a
qualitatively similar result for other microscopic models.
From the limited spatial extent of the Kondo resonance for Ce on
Ag(111)~\cite{STM98_ce} and Co on Au(111)~\cite{STM98_co}
we thus conclude that $q_0$ is effectively zero above a
lateral distance of about 10\AA\ in these experiments.

Our calculations further indicate that the spatial dependence
of the differential conductance as measured for Co on Au(111)
is not generic, but intimately depends on the microscopic
details of Au(111). Indeed, while Madhavan {\em et al.} observe a
dip-like structure that persists up to a lateral distance of
$\sim$10\AA\ from the adatom,~\cite{STM98_co} we typically
find a resonance at a distance of one lattice spacing. In
Fig.~\ref{fig:fig6}, where point tunneling is assumed, this
resonance occurs for any value of $q_0$, which is a special
feature of the half-filled nearest-neighbor tight-binding model
used. One may anticipate, though, a similar resonance within
a range of one lattice spacing for other lattice models
at half filling, since $G_{\vec{R}_s,\vec{R}_i}(0+i\eta)$ is
expected to oscillate on a length scale of one
lattice spacing. We emphasize, however,
that the spatial dependence measured by Madhavan {\em et al.}
remains quite restrictive for the Hamiltonian of
Eq.~(\ref{H_tight-binding}) both away from half filling and
in the case of nonlocal tunneling between the tip and the
substrate conduction electrons.

The particular tight-binding model used in this paper clearly
limits the application of our results to the experiments. To
make direct contact with the experimental data it is necessary
to employ realistic Green functions for the (111) surfaces of
silver and gold, which may be obtained, for example, from {\em ab-inito}
calculations. It would be interesting to see if the combination
of {\em ab-inito} calculations for the (111) surface of gold
with NCA calculations for the many-body Kondo resonance can
reproduce the particular spatial dependence of the
differential conductance as seen for Co on Au(111).

Another interesting issue is the magnetic-field dependence of
the low-temperature, low-voltage differential conductance.
With increasing magnetic field, the Abrikosov-Suhl resonance
is first split for a moderate magnetic field, $H \sim T_K$, before
a large magnetic field suppresses the Kondo effect altogether.
A similar pattern is expected for the Kondo resonance in the
differential conductance.
Unfortunately, treatment of a finite magnetic field within the
NCA is hampered by the NCA pathology,~\cite{Bickers87} hence
a different approach is required. One possibility might be Quantum
Monte Carlo simulations in combination with the Maximum Entropy
method for analytic continuation,~\cite{Amit-private_comm}
although such an approach is restricted in treating realistically
small Kondo temperatures.

Finally, in this paper we have focused on the case of an
individual magnetic adatom; however, using the STM tip to
atomically manipulate individual adatoms into forming
small clusters, it might be possible to address
the subtle interplay between the Kondo effect and magnetic
correlations among the different adatoms. Most notably, the
competition between the Kondo effect and antiferromagnetic
locking in the case of two close-by adatoms.~\cite{two_imp}
A first study of a Co dimer on Au(111) along these lines
was recently reported in Ref.~\onlinecite{Co_dimer}.
As detailed in Appendix~\ref{appendix:multiple_adatoms},
our formulation of the tunneling current is naturally
extended to the case of multiple magnetic adatoms. Specifically,
the single $d$-electron Green function entering Eq.~(\ref{G^f})
is replaced by a matrix propagator, corresponding to all
possible propagations within the adatom cluster. In this
manner, one can analyze complicated multiple-adatom
configurations in terms of the intra-site and
inter-site $d$ Green functions.

It is our hope that the approach developed in this paper
will prove useful in analyzing future STM measurements
of magnetic adatoms on metallic surfaces.

\section*{Acknowledgements}
A. S. is grateful to Wilfrid Aulbur, Frithjof Anders, Amit Chattopadhyay,
Daniel Cox, and John Wilkins for stimulating discussions. S. H. was
supported in part by NSF grant DMR9357474, the NHMFL, and the Research
Corporation.

\appendix
\section{Simplified expressions for the first three nonlocal surface
Green functions}
\label{appendix:surface_GF}
In this appendix, we provide simplified expressions for the nonlocal surface
Green functions $G_{01}$, $G_{11}$, and $G_{02}$, involving just a single
integration. These expressions are analogous to Eq.~(\ref{concise_G_00}) for
$G_{00}$.

Denoting for convenience $z = \epsilon + i\eta$ and
$\zeta_k = z - 2t\cos(k)$, the nonlocal
Green functions are expressed as
\begin{equation}
G_{lm}(z) = \int_{0}^{\pi}\!\!
            G_{2D}(\zeta_k) F_{lm}(z,k) \frac{dk}{\pi} ,
\end{equation}
with
\begin{equation}
F_{01}(z, k) = \cos(k) - \cos(2k)\frac{\zeta_k}{4t},
\end{equation}
\begin{eqnarray}
F_{11}(z, k) &=& \cos(k) \left \{ \left( 2 - \frac{z^2}{6t^2}\right)
            \frac{\zeta_k}{4t} + \frac{z}{3t} \right\}
\nonumber\\
&+& \cos(2k) \frac{z\zeta_k}{12t^2} ,
\end{eqnarray}
and
\begin{eqnarray}
F_{02}(z, k) &=& \cos(k) \left \{ \left( \frac{z^2}{3t^2} - 1\right)
            \frac{\zeta_k}{4t} - \frac{2z}{3t} \right\}
\nonumber\\
&+& \cos(2k) \left \{ 2 - \frac{5z\zeta_k}{12t^2} \right \}
\nonumber\\
&+& \cos(3k) \frac{\zeta_k}{4t} .
\end{eqnarray}
Similar expressions, but with modified $F_{lm}(z, k)$, apply
also to $G_{lm}(z)$ with larger values of $|l| + |m|$.

\section{Comparison with previous work}
\label{appendix:comparison}
It is instructive to compare the present theory of point tunneling
between the STM tip and the substrate conduction electrons,
Eqs.~(\ref{G^f})--(\ref{differential_cond}), with the analyses of
Refs.~\onlinecite{STM98_ce} and \onlinecite{STM98_co}, which
focused on the case where $\vec{R}_i = \vec{R}_s$. In
Ref.~\onlinecite{STM98_ce}, Li {\em et al.} considered the case of zero
direct tunneling between the tip and adatom, corresponding to $q = 0$
in Fano's notation. The antiresonance that develops in $G(V)$
in this case was approximated by the inverted NCA lineshape
of the impurity spectral function, computed within a degenerate
Anderson model that accounts for the full $4f$ degeneracy in Ce.
The depth of the antiresonance was left as a fitting parameter.

As evident from Eqs.~(\ref{G^f})--(\ref{differential_cond})
with $\vec{R}_i = \vec{R}_s$, the above relation between the
impurity contribution to the differential conductance and the
impurity spectral function is exact in the limit $T \to 0$,
provided $G_{00}(\epsilon + i\eta) = -i \pi \rho_0$ is essentially
constant for $|\epsilon|$ on the scale of $T_K$. This relation loses
accuracy, though, for $T > T_K$, when the convolution with the derivative
of the Fermi-Dirac function in Eq.~(\ref{differential_cond})
increasingly smears the lineshape of the impurity spectral function.

Contrary to Li {\em et al.}, who restricted attention to $q = 0$,
Madhavan {\em et al.} considered the full range in $q$. To this end,
Fano's expression for the differential conductance was generalized
according to~\cite{STM98_co}
\begin{equation}
G(V) = G_0 \frac{(q + eV')^2}{1 + (eV')^2} ,
\label{Crommie-Fano_1}
\end{equation}
\begin{equation}
eV' = \frac{ eV - \epsilon_d - {\rm Re}
          \left \{ \Sigma_{\sigma}^d(eV + i\eta) \right \} }
          { {\rm Im}\left\{ \Sigma_{\sigma}^d(eV + i\eta) \right \} } ,
\label{Crommie-Fano_2}
\end{equation}
where $\Sigma_{\sigma}^d(\epsilon + i\eta)$ is the full $d$-electron
self-energy,
including both the on-site repulsion $U$ and the hybridization to the
conduction band, $V_h$. The $d$ self-energy was approximated in turn by
a form corresponding to a Lorentzian Abrikosov-Suhl resonance, with a
half-width $T_K$ and a peak position that was left as a fitting parameter.

Comparison with Eqs.~(\ref{G^f})--(\ref{differential_cond})
for $\vec{R}_i = \vec{R}_s$ reveals that
Eqs.~(\ref{Crommie-Fano_1})--(\ref{Crommie-Fano_2}) are correct in
the limit $T \to 0$, provided
${\rm Im}\left \{\Sigma_{\sigma}^d(\epsilon + i\eta) \right \} =
-\Gamma$. The latter equality is exact for $T = 0$ and
$\epsilon = 0$,~\cite{Langreth66} and is a reasonable approximation
for $T < T_K$ and $|\epsilon| \alt T_K$.
This approach, however, breaks down for $T > T_K$,
both due to the inapplicability of the assumed form of
$\Sigma_{\sigma}^d(\epsilon + i\eta)$, and because of the convolution
with the derivative of
the Fermi-Dirac function in Eq.~(\ref{differential_cond}) which smears
the underlying structure of $G_{f\sigma}(\epsilon + i\eta)$. Hence,
similar to the analysis of Li {\em et al.}, this approach is restricted
to the low-temperature regime. By contrast,
Eqs.~(\ref{G^f})--(\ref{differential_cond}) are valid for any temperature
$T$ and any $\vec{R}_s \neq \vec{R}_i$, and are easily amendable [using
Eqs.~(\ref{sub_G's})] to the case of nonlocal tunneling
between the STM tip and the substrate conduction electrons.

\section{Several magnetic adatoms}
\label{appendix:multiple_adatoms}
In this appendix, we generalize our formulation of the tunneling
current to the case of several magnetic adatoms deposited on top
of the metallic surface. Specifically, we consider a cluster of
$m$ close-by adatoms positioned at points $\vec{R}_j$ ($j = 1,
\cdots, m$), each with its own hybridization matrix element,
$V_{hj}$, and its own tunneling matrix element, $t_j$. The
different adatoms need not be identical, and can generally
have different $d$-level energies and different on-site Coulomb
repulsions (denoted by $\epsilon_j$ and $U_j$, respectively).
The Hamiltonian of the system has the form
${\cal H} = {\cal H}_{sub} + {\cal H}_{tip} + {\cal H}_{tun}$,
where ${\cal H}_{tip}$ is described by Eq.~(\ref{H_tip}),
and ${\cal H}_{sub}$ and ${\cal H}_{tun}$ are given by
\begin{eqnarray}
{\cal H}_{sub} &=& \sum_{\vec{k}\sigma} \epsilon_{\vec{k}}
           c^{\dagger}_{\vec{k}\sigma}c_{\vec{k}\sigma}
           + \sum_j \left \{ \epsilon_{j}
             \sum_{\sigma} n^{d}_{j\sigma} + U_j n^{d}_{j\uparrow}
             n^{d}_{j\downarrow} \right\}\nonumber\\
&+& \sum_{j,\sigma} V_{h j}
           \left \{ d^{\dagger}_{j\sigma} \psi_{\sigma}(\vec{R}_j) +
           \psi^{\dagger}_{\sigma}(\vec{R}_j)d_{j\sigma} \right \} ,
\label{H_sub_mi}\\
{\cal H}_{tun} &=& 
           t_c\sum_{\sigma} \left \{ 
           \psi^{\dagger}_{\sigma}(\vec{R}_s) A_{\sigma} +
           A^{\dagger}_{\sigma}\psi_{\sigma}(\vec{R}_s) \right \}
\nonumber\\
&+& \sum_{j}
           t_{j} \sum_{\sigma} \left \{ d^{\dagger}_{j\sigma}A_{\sigma} +
           A^{\dagger}_{\sigma}d_{j\sigma} \right \} .
\label{H_tun_mi}
\end{eqnarray}
Here $d^{\dagger}_{j\sigma}$ creates an atomic $d$ electron
with spin $\sigma$ on the $j$-th adatom, and
$n^{d}_{j\sigma} = d^{\dagger}_{j\sigma}d_{j\sigma}$ is the
corresponding number operator. All other notations are the
same as in Eqs.~(\ref{H_sub})--(\ref{H_tun}).

Evaluating the tunneling current from the substrate to the tip
to second order in $t_c$ and $t_j$, one obtains an expression
identical to that of Eq.~(\ref{current}), with the sole
modification that $\rho_{f\sigma}(\epsilon) =
-{\rm Im}G_{f\sigma}(\epsilon + i\eta)$ represents
now the zero-tunneling spectral function corresponding to
\begin{equation}
f_{\sigma} = t_c \psi_{\sigma}(\vec{R}_s)
           + \sum_j t_j d_{j\sigma} .
\end{equation}
Introducing the $m\!\times\!m$ matrix $d$ Green function
\begin{eqnarray}
&&G^{d}_{ij\sigma}(\epsilon + i\eta) = \int_{-\infty}^{\infty}
         G^{d}_{ij\sigma}(t, t') e^{i\epsilon (t-t')} dt ,
\label{matrix_d_GF} \\
&&G^{d}_{ij\sigma}(t, t') = -\theta(t - t') \langle \{ d_{i\sigma}(t),
         d^{\dagger}_{j\sigma}(t') \}\rangle ,
\end{eqnarray}
together with the two ``vector'' quantities
\begin{eqnarray}
v_j(\epsilon + i\eta) &=&
      t_j + t_c V_{hj} G_{\vec{R}_s, \vec{R}_j}(\epsilon + i\eta) ,\\
u_j(\epsilon + i\eta) &=&
      t_j + t_c V_{hj} G_{\vec{R}_j, \vec{R}_s}(\epsilon + i\eta) ,
\end{eqnarray}
the retarded $f$ Green function is conveniently expressed as
\begin{eqnarray}
G_{f\sigma}(\epsilon + i\eta) &=& 
      t_c^2 G_{\vec{R}_s, \vec{R}_s}(\epsilon + i\eta) \nonumber\\
&+& \sum_{i,j} v_i(\epsilon + i\eta)
      G_{ij\sigma}^{d}(\epsilon + i\eta)
      u_j(\epsilon + i\eta) ,
\label{G_f_cluster}
\end{eqnarray}
which has the compact matrix representation:
\begin{equation}
G_{f\sigma}(\epsilon + i\eta) =
      t_c^2 G_{\vec{R}_s, \vec{R}_s}(\epsilon + i\eta)
      + \left[ v^{t}G^{d}_{\sigma} u \right] (\epsilon + i\eta) .
\end{equation}

All information of the adatom cluster and its many-body
physics is contained within the $G_{f\sigma}$ Green
function of Eq.~(\ref{G_f_cluster}), which replaces
that of Eq.~(\ref{G^f}) in the final expression for the
differential conductance, Eq.~(\ref{differential_cond}).
In particular, Eq.~(\ref{G^f}) is properly recovered in
the case of just a single magnetic adatom. This permits
the analysis of complicated multiple-adatom configurations
in terms of the matrix $d$ Green function of
Eq.~(\ref{matrix_d_GF}).

\end{document}